\documentclass[aps,prl,twocolumn,showpacs,preprintnumbers,amsmath,amssymb,superscriptaddress]{revtex4}%
\usepackage{graphicx}%
\usepackage{dcolumn}
\usepackage{amsmath}

\makeatletter
\def\btt#1{\texttt{\@backslashchar#1}}%
\DeclareRobustCommand\bblash{\btt{\@backslashchar}}%
\makeatother

\begin{document}
%
%
\draft

\title{Pressure effect on the
in-plane magnetic penetration depth in YBa$_2$Cu$_4$O$_{8}$}
\author{R.~Khasanov}
\affiliation{ Laboratory for Neutron Scattering, ETH Z\"urich and
Paul Scherrer Institut, CH-5232 Villigen PSI, Switzerland}
\affiliation{DPMC, Universit\'e de Gen\`eve, 24 Quai
Ernest-Ansermet, 1211 Gen\`eve 4, Switzerland}
\affiliation{Physik-Institut der Universit\"{a}t Z\"{u}rich,
Winterthurerstrasse 190, CH-8057, Z\"urich, Switzerland}
\author{J.~Karpinski}
\affiliation{Solid State Physics Laboratory, ETH 8093 Z\"urich,
Switzerland}
\author{H.~Keller}
\affiliation{Physik-Institut der Universit\"{a}t Z\"{u}rich,
Winterthurerstrasse 190, CH-8057, Z\"urich, Switzerland}

%
\begin{abstract}
We report a study of the pressure effect (PE) on the in-plane
magnetic field penetration depth $\lambda_{ab}$ in
YBa$_2$Cu$_4$O$_{8}$~ by means of Meissner fraction measurements.
A pronounced PE on $\lambda_{ab}^{-2}(0)$ was observed with a
maximum relative shift of
$\Delta\lambda^{-2}_{ab}/\lambda^{-2}_{ab}= 44(3)\%$ at a pressure
of 10.2~kbar. It arises from the pressure dependence of the
effective in-plane charge carrier mass and pressure induced charge
carrier transfer from the CuO chains to the superconducting
CuO$_2$ planes.  The present results imply that the charge
carriers in YBa$_2$Cu$_4$O$_{8}$ are coupled to the lattice.

\end{abstract}
\maketitle
%

One of the fundamental questions concerning the physics of the
cuprate high-temperature superconductors (HTS) is whether the
electron-phonon interaction plays an essential role in these
systems or not. The conventional phonon-mediated theory of
superconductivity is based on the Migdal adiabatic approximation
in which the effective supercarrier mass $m^{\ast}$ is independent
of the lattice vibrations. However, if the interaction between the
carriers and the lattice is strong enough, the Migdal adiabatic
approximation breaks down and $m^{\ast}$ depends on the lattice
degrees on freedom (see, {\it e.g.} \cite{Alexandrov94}). One way
to explore a possible coupling of the supercarriers to the lattice
is a study how the magnetic field penetration depth $\lambda$ is
affected by the crystal lattice modifications. For cuprate
superconductors (clean limit) the zero temperature in-plane
penetration depth $\lambda_{ab}(0)$ is proportional to the
superfluid density $n_{s}/m_{ab}^{\ast}$ ($n_{s}$ is the
superconducting charge carrier density and $m_{ab}^{\ast}$ is the
in-plane effective mass of the superconducting charge carriers):
\begin{equation}
\lambda_{ab}^{-2}(0) \propto n_{s}/m_{ab}^{\ast}.
 \label{eq:lambda}
\end{equation}
Under the assumption that one can separate the two quantities
$n_s$ and $m^\ast$, Eq.~(\ref{eq:lambda}) implies that a change of
$\lambda_{ab}$ is due to a shift in $n_{s}$ and/or
$m_{ab}^{\ast}$:
\begin{equation}
\Delta\lambda^{-2}_{ab}(0)/\lambda^{-2}_{ab}(0)= \Delta n_s/n_s
-\Delta m_{ab}^{\ast}/m_{ab}^{\ast}.
 \label{eq:Deltalambda}
\end{equation}
Therefore, if the contribution of $n_{s}$ is known,  $\lambda$
measurements open an unique possibility to investigate an unusual
(non-adiabatic) coupling of the charge carriers to the crystal
lattice in HTS.

One way to explore the role of lattice vibrations in HTS is to
perform the isotope effect experiments. Previous studies
\cite{Zhao97,Zhao98,Zhao01,Hofer00,Khasanov03,Khasanov03b} showed
a substantial oxygen-isotope ($^{16}$O/$^{18}$O) effect  on the
in-plane penetration depth $\lambda _{ab}$, which indicates a
non-adiabatic coupling of the electrons to phonon modes involving
the movement of the isotope substituted atoms. An alternative way
to explore lattice effects in HTS is pressure experiments. The
squeezing of the crystal lattice by external hydrostatic or
uniaxial pressure affects the lattice parameters, the phonon
spectrum and consequently the electron-lattice coupling.
Surprisingly, the pressure effect (PE) on the magnetic field
penetration depth has not attracted much attention. To our
knowledge only one experiment related to this topic for the
two-dimensional organic superconductor $\kappa - ({\rm
BEDT-TTF})_2 {\rm Cu(NCS)}_2$ was reported so far \cite{Larkin01}.
It was shown that both $\lambda^{-2}(0)$ and the transition
temperature $T_c$ linearly decreases with pressure $p$.
Furthermore,  $T_c$ vs. $\lambda^{-2}(0)$ measured for different
$p$ follows the universal "Uemura" line \cite{Uemura89,Uemura91}.

In this letter we report the first investigation of the in-plane
magnetic penetration depth $\lambda_{ab}$ under high hydrostatic
pressure (up to 10.2~kbar) in YBa$_2$Cu$_4$O$_{8}$ (Y124). The
temperature dependence of $\lambda^{-2}_{ab}$ was extracted from
Meissner fraction measurements at low magnetic field. A pronounced
pressure effect on $\lambda_{ab}$ with a maximum relative shift
$\Delta\lambda^{-2}_{ab}(0)/\lambda^{-2}_{ab}(0)=44(3)\%$ at a
hydrostatic pressure $p=$10.2~kbar was observed. We demonstrate
that this effect mainly ($\simeq 2/3$) arises from the pressure
dependence of the in-plane charge carrier mass $m^\ast_{ab}$.

\begin{figure}[htb]
\includegraphics[width=1.0\linewidth]{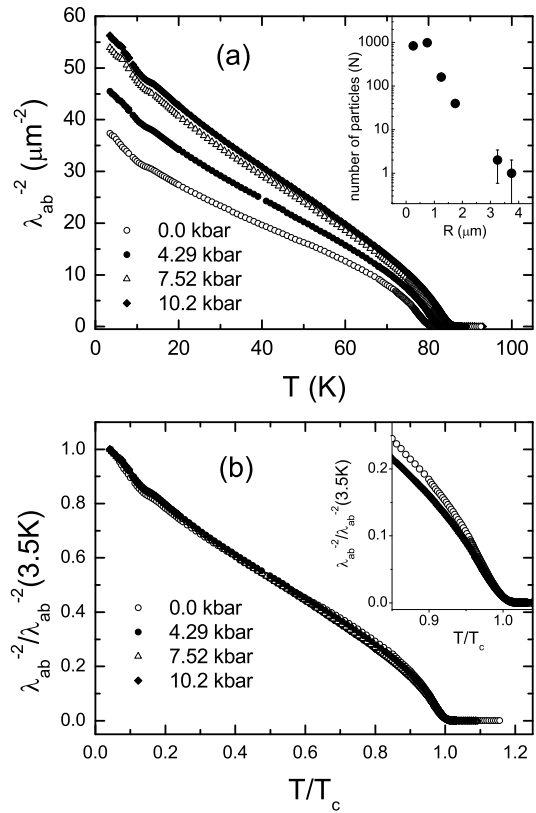}
\caption{  (a) $\lambda^{-2}_{ab}$ as a function of $T$  for
YBa$_2$Cu$_4$O$_{8}$\ fine powder  sample at hydrostatic pressures
$p=$~0.0, 4.29, 7.52 and 10.2~kbar as obtained from low-field
SQUID magnetization measurements.  Inset shows the grain size
distribution in a semilogarithmic scale. Errors are statistical.
(b) Normalized superfluid density
$\lambda^{-2}_{ab}(T)/\lambda^{-2}_{ab}(3.5$~K) as a function of
reduced temperature $T/T_c$ for the same pressures. The insert
shows the data for $p=$~0.0~kbar and 10.2~kbar close to $T/T_c=1$.
}
\label{fig:lambda_vs_T}
\end{figure}

The polycrisatalline YBa$_2$Cu$_4$O$_{8}$ samples were synthesized
by solid-state reactions using high-purity Y$_2$O$_3$, BaCO$_3$
and CuO. The powder samples were ground for about 60~min and then
passed through the 10~$\mu$m sieve in order to obtain very small
grains that are needed for the determination of $\lambda$ from
Meissner fraction measurements.
The grain size distribution of the powder was then determined by
analyzing SEM (scanning electron microscope) photographs. The
measured particle radius distribution $N(R)$ is shown in the inset
of Fig.~\ref{fig:lambda_vs_T}(a).
The hydrostatic pressure was generated in a copper-beryllium
piston cylinder clamp that was especially designed for
magnetization measurements under pressure \cite{Straessle02}. The
sample was mixed with Fluorient FC77 (pressure transmitting
medium) with a sample to liquid volume ratio of approximately
$1/6$.  With this cell hydrostatic pressures up to 12~kbar can be
achieved \cite{Straessle02}. The pressure was measured in situ  by
monitoring the $T_{c}$ shift of the small piece of In
[$T_{c}(p=0)\simeq 3.4$~K] included in the pressure cell. The
value of the Meissner fraction $f$ was calculated from 0.5~mT
field-cooled (FC) SQUID magnetization measurements. The absence of
weak links between the grains was confirmed by the linear magnetic
field dependence of the FC magnetization in low fields (0.25, 0.5,
0.75, and 10~mT) measured at 8~K for each pressure (0.0, 4.29,
7.52, and 10.2~kbar).

The temperature dependence of $\lambda_{\rm eff}$ (powder average)
was calculated from $f$ by using the Shoenberg model
\cite{Shoenberg40} modified for the known grain size distribution
\cite{Porch93}.
The in-plane penetration depth $\lambda_{ab}(T)$
[Fig.~\ref{fig:lambda_vs_T}(a)] was determined from the measured
$\lambda_{\rm eff}(T)$ using the relation $\lambda_{\rm
eff}=1.31\lambda_{ab}$, which holds for highly anisotropic
superconductors ($\lambda_c/\lambda_{ab}>5$) \cite{Fesenko91}.
The values of $T_c$ and $\lambda_{ab}^{-2}(0)$ at each particular
pressure were defined as: $T_c$ -- from the intersect of the
linearly extrapolated $\lambda_{ab}^{-2}(T)$  in the vicinity of
the superconducting transition with the $\lambda_{ab}^{-2}=0$
line; $\lambda_{ab}^{-2}(0)$ --  from the intersect of the linear
fit of $\lambda_{ab}^{-2}(T)$ at $T<10$~K with the $T=0$ line.
Note that the value of $\lambda_{ab}(0)$ at ambient pressure was
found to be $\lambda_{ab}(0)=156$~nm in a good agreement with the
literature data \cite{Keller91,Bernhard95}. The values of $\lambda
_{ab}^{-2}(0)$ and $T_{c}$ obtained from the experimental data
presented in Fig.~\ref{fig:lambda_vs_T} are summarized in Table
~\ref{Table1}. The errors in $\lambda_{ab}^{-2}(0)$ come from
misalignments of the experimental setup after the cell was removed
from the SQUID magnetometer and put back again. We checked this
procedure with a set of measurements at constant pressure. The
systematic scattering of the magnetization data is of about
$0.5\%$, giving a relative error in $\lambda^{-2}(T)$ of about
2\%.

Figure~\ref{fig:lambda_vs_T}(b) shows a plot of the normalized
superfluid density $\lambda _{ab}^{-2}(T)/\lambda
_{ab}^{-2}(3.5$K) versus the reduced temperature $T/T_{c}$. For
all pressures the temperature dependencies of $\lambda _{ab}^{-2}$
are nearly the same. At low temperatures ($T/T_{c}<0.5$) all
curves appear to collapse, while by approaching $T_{c}$
($0.6<T/T_{c}<1.0$) they begin to separate. As it is shown in the
inset of Fig.~\ref{fig:lambda_vs_T}(b) $\lambda
_{ab}^{-2}(T)/\lambda _{ab}^{-2}(3.5$K)  decreases with increasing
pressure. Note, that $\lambda^{-2}_{ab}(T)$ curves of the present
work  are very similar  to those reported in
Ref.~[\onlinecite{Panagopulos99}] obtained for
YBa$_2$Cu$_4$O$_{8}$ using the same experimental technique. The
fast increase of $\lambda^{-2}_{ab}$ close to $T_c$ (see
Fig.~\ref{fig:lambda_vs_T}) in Ref.~[\onlinecite{Panagopulos99}]
was attributed to the contribution  of the double chains to
$\lambda^{-2}_{ab}$. Thus the decreasing of $\lambda
_{ab}^{-2}(T)/\lambda _{ab}^{-2}(3.5$K) close to $T/T_{c}=1$ [see
inset in Fig.~\ref{fig:lambda_vs_T}(b)] would imply that the
contribution of the chains to $\lambda_{ab}^{-2}$ {\it decreases}
with increasing pressure.

\begin{figure}[htb]
\includegraphics[width=1.0\linewidth]{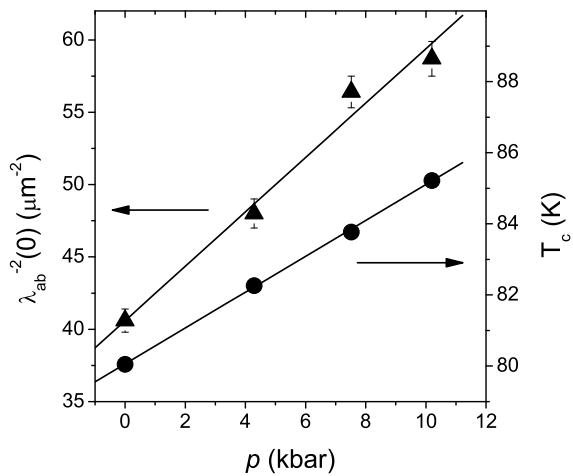}
\caption{  Pressure dependence of the transition temperature $T_c$
($\bullet$) and the zero temperature in-plane magnetic penetration
depth $\lambda^{-2}_{ab}(0)$ ($\blacktriangle$). The linear fits
yield ${\rm d}T_c/{\rm d}p = 0.50(1)\ {\rm K/kbar}$ and ${\rm
d}\lambda^{-2}_{ab}(0)/{\rm d}p = 1.88(13)\ {\rm \mu
m^{-2}/kbar}$. The errors in $\lambda_{ab}^{-2}(0)$ come from
misalignments of the experimental setup (see text for an
explanation).  }
\label{TcLambda}
\end{figure}

Fig.~\ref{TcLambda} shows the pressure dependence of $T_c$ and
$\lambda^{-2}_{ab}(0)$ extracted form the data presented in
Fig.~\ref{fig:lambda_vs_T}. As it is seen $\lambda_{ab}^{-2}(0)$
as well as $T_c$ increase with increasing pressure almost
linearly. The linear fit yield ${\rm d}T_c/{\rm d}p = 0.50(1)\
{\rm K/kbar}$ and ${\rm d}\lambda^{-2}_{ab}(0)/{\rm d}p =
1.88(13)\ {\rm \mu m^{-2}/kbar}$ for $T_c(p)$ and
$\lambda^{-2}_{ab}$, respectively. The value of $dT_c/dp$ is in
good agreement with the literature data (see e.g.
\cite{Wijngaarden99}). The values of the relative pressure shifts
$\Delta T_c/T_c$ and $\Delta
\lambda_{ab}^{-2}(0)/\lambda_{ab}^{-2}(0)$ are summarized in
Table~\ref{Table1}. The pressure shift of a physical quantity $X$
is defined as: $\Delta X/X=[X_{p>0}-X_{p=0}]/X_{p=0}$.

\begin{table}
 \caption[~]{Summary of the PE results for
YBa$_2$Cu$_4$O$_{8}$~ extracted from the experimental data (see
text for an explanation). }
 \label{Table1} %
\begin{center}
\begin{tabular}{cccccccc} \hline\hline

\hline
p& $T_c$  & $\frac{\Delta T_c}{T_c}$ &
$\lambda^{-2}_{ab}(0)$ &$\frac{\Delta
\lambda_{ab}^{-2}(0)}{\lambda_{ab}^{-2}(0)}$&$\frac{\Delta
m^\ast_{ab}}{m^\ast_{ab}}$ \\
(kbar)&(K)&(\%)&($\mu m^{-2}$)&(\%)&(\%)\\
%
 \hline
0.0&80.05(2)&-&40.6(8)&-&- \\
4.29&82.26(2)&2.8(4)&48.0(1.0)&18(3)&-13(2) \\
7.52&83.77(3)&4.1(5)&56.4(1.1)&39(3)& -29(3)\\
10.2&85.22(3)&6.5(5)&58.7(1.2)&44(3)&-32(3)\\
\hline \hline
\end{tabular}
\end{center}
\end{table}

\begin{figure}[htb]
\includegraphics[width=1.0\linewidth]{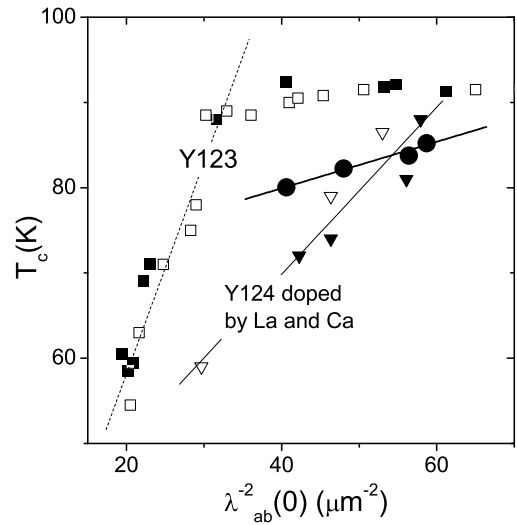}
\caption{$T_c$ versus $\lambda^{-2}_{ab}(0)$ for various Y123 and
Y124 samples. The dashed line corresponds to the  ''Uemura`` line
\cite{Uemura89,Uemura91}. The open and closed squares are Y123
data from Zimmermann {\em et al.} \cite{Zimmermann95} and Tallon
{\em et al.} \cite{Tallon95}. Down and up open triangles are data
of Y$_{1-x}$Ca$_x$Ba$_2$Cu$_4$O$_8$ and
YBa$_{2-x}$La$_x$Cu$_4$O$_8$ taken from Shengelaya {\em et al.}
\cite{Shengelaya98} and Tallon {\em et al.} \cite{Tallon95}.
Closed circles are the pressure data of YBa$_2$Cu$_4$O$_{8}$ from
the present study. The solid  lines are linear fits of $T_c$ vs
$\lambda^{-2}_{ab}(0)$. While the slope of the linear fit to the
doped samples is dominated by changes on $n_{s}$ only, the slope
of the YBa$_2$Cu$_4$O$_{8}$\ pressure data is determined mostly
($\sim70$~\%) by changes in $m^\ast_{ab}$ (see text for an
explanation). }
\label{uemura_plot}
\end{figure}

Now we are turning to discuss the zero-temperature values of
$\lambda^{-2}_{ab}(0)$. Fig.~\ref{uemura_plot} shows $T_c$ plotted
versus $\lambda^{-2}_{ab}(0)$  for our YBa$_2$Cu$_4$O$_{8}$~
sample at different pressures (see Table~\ref{Table1}), together
with previous results of Y$_{1-x}$Ca$_x$Ba$_2$Cu$_4$O$_8$ and
YBa$_{2-x}$La$_x$Cu$_4$O$_8$  \cite{Tallon95, Shengelaya98} and
oxygen deficient YBa$_2$Cu$_3$O$_{7-\delta}$ (Y123)
\cite{Tallon95, Zimmermann95}.
We used the relation $\lambda^{-2}_{ab}(0)(\mu
m^{-2})=\sigma(0)(\mu s^{-1})/(0.266)^2$ \cite{Niedermayer02} to
calculate $\lambda^{-2}_{ab}(0)$ from the values of the
zero-temperature muon-spin depolarization rate $\sigma(0)$
reported in Refs.~[\onlinecite{Tallon95,
Shengelaya98,Zimmermann95}].
In the underdoped regime $T_c$ scales linearly with
$\lambda^{-2}_{ab}(0)$ on a single universal line for most HTS
families (dashed line in Fig.~\ref{uemura_plot}). This is a
generic behavior expected for HTS that contain CuO$_2$ planes only
\cite{Uemura89,Uemura91}. It was found that several HTS systems
containing CuO chains exhibit enhanced values of
$\lambda^{-2}_{ab}(0)$ compared to the "Uemura line"
\cite{Tallon95}. This deviation from the Uemura scaling was
explained by an additional contribution to $\lambda^{-2}_{ab}(0)$
from the disorder free CuO chains \cite{Tallon95}.

Under the assumption that Eq.~(\ref{eq:Deltalambda}) is valid,
there are two contributions which may cause a change of
$\lambda_{ab}(0)$ under pressure: (i) the charge carrier
concentration $n_s$ and (ii) the in-plane charge carrier mass
$m^\ast_{ab}$. In order to separate them, we use the following
procedure. Doping of the parent compound YBa$_2$Cu$_4$O$_{8}$~
with La or Ca \cite{Bernhard95,Tallon95,Shengelaya98} leads mainly
to a change of the charge carrier concentration $n_s$ within
superconducting CuO$_2$ planes \cite{Sakurai90,Chatterjee99}. In
addition, the change of the lattice parameters upon La or Ca
doping in Y124 is negligibly small \cite{Sakurai90}. Therefore,
doping does not affect $m^{\ast}_{ab}$ very much. Combining these
two arguments one can conclude that changes in $\lambda_{ab}$ due
to doping are determined mainly by $n_{s}$, and
Eq.~(\ref{eq:Deltalambda}) yields:
\begin{equation}
\left.\frac{\Delta\lambda_{ab}^{-2}}{\lambda_{ab}^{-2}}\right|_d\simeq
\frac{\Delta n_{s}}{n_{s}}
\label{eq:DelataLambdaDoping}
\end{equation}
Here the index $d$ stays for doping.

The PE on $T_{c}$ observed in HTS is usually explained in terms of
the so-called pressure-induced charge transfer model
\cite{Wijngaarden99,Almasan92,Neumeier93,Gupta95}. In this model
there are two pressure dependent parameters, one is the charge
transfer to the superconducting CuO$_{2}$ planes, and the second
is the maximum value of transition temperature at optimal doping
[$T_{c}^{max}(p)$]. In underdoped YBa$_{2}$Cu$_{4}$O$_{8}$,
pressure induces a charge carrier transfer from the chains to the
planes, and ${\rm d}T_{c}^{max}(p)/{\rm d}p$ is negligibly small
compared to the charge transfer term \cite{Wijngaarden99}. This
finding is further supported by NQR measurements under pressure
\cite{Zimmermann90} and by our $\lambda _{ab}^{-2}(T)$ data,
indicating that the contributions from the chains to $\lambda
_{ab}^{-2}$ decreases with increasing pressure [see inset in
Fig.~\ref{fig:lambda_vs_T}~(b)]. Moreover, pressure changes the
lattice parameters and, consequently, may affect $n_s$ and
$m^\ast_{ab}$. Therefore, in Eq.~(\ref{eq:Deltalambda}) both terms
are pressure dependent. Furthermore, in HTS cuprates $T_c$ is a
universal function of doping. Assuming that the shift of $T_c$ due
to doping and pressure is caused by the same change of $n_{s}$
($\left.\Delta n_s/n_s\right|_d=\left.\Delta n_s/n_s\right|_p$),
then from Eqs.~(\ref{eq:Deltalambda}) and
(\ref{eq:DelataLambdaDoping}) one readily obtains:
\begin{equation}
\left.\frac{\Delta m^\ast_{ab}}{ m^\ast_{ab}}\right|_p \simeq
\left.\frac{\Delta \lambda_{ab}^{-2}}{\lambda_{ab}^{-2}}\right|_d
- \left.\frac{\Delta
\lambda_{ab}^{-2}}{\lambda_{ab}^{-2}}\right|_p.
\label{eq:DeltaLambdaPressure}
\end{equation}
Here the index  $p$ means pressure. Linear fits of
$\lambda^{-2}_{ab}(0)$ vs. $T_c$ (see Fig.~\ref{uemura_plot})
give: $\left. \Delta \lambda_{ab}^{-2}(0)\right|_d=1.0(1)\cdot
\Delta T_c$ and $
\left.\Delta\lambda_{ab}^{-2}(0)\right|_p=3.7(3)\cdot \Delta T_c$
for La and Ca doped Y124 and for YBa$_2$Cu$_4$O$_{8}$ under
pressure, respectively. Thus the pressure dependence of the
in-plane charge carrier mass may be written as
\begin{equation}
\left.\frac{\Delta m^\ast_{ab}}{m^\ast_{ab}}\right|_p \simeq
-0.72(6)\times
\left.\frac{\Delta\lambda_{ab}^{-2}(0)}{\lambda_{ab}^{-2}(0)}\right|_p.
 \label{eq:mass}
\end{equation}
The values of $\left.\Delta m^\ast_{ab}/ m^\ast_{ab}\right|_p$
obtained from Eq.~(\ref{eq:mass}) are summarized in
Table~\ref{Table1}.

In summary, we report the first observation of the pressure effect
on the zero temperature in-plane magnetic field penetration depth
$\lambda_{ab}(0)$ in a cuprate superconductor. A pronounced PE on
both the transition temperature $T_c$ and $\lambda_{ab}^{-2}(0)$
is observed which increases with increasing pressure. The pressure
shift on $\lambda_{ab}^{-2}(0)$ is attributed to (i) the pressure
induced charge carrier transfer from chains to the planes and (ii)
the decreasing of the in-plane charge carrier mass $m^\ast_{ab}$.
At $p=$10.2~kbar we observed $\Delta
\lambda_{ab}^{-2}(0)/\lambda_{ab}^{-2}(0) = 44(3)\%$ and $\Delta
m_{ab}^{\ast}/m_{ab}^{\ast} = -32(3)\%$. Such a large effect on
$m_{ab}^{\ast}$ implies that lattice effects play an essential
role in cuprate  superconductors.

We are grateful to D.Di~Castro, D.G.~Eshchenko, H.~Luetkens, and
J.~Roos for fruitful discussions, K.~Conder for help during sample
preparation,  T.~Str\"assle for designing the pressure cell for
magnetization measurements, and R.~Br\"utsch and D.~Gavillet for
the measurements of the grain size distribution. This work was
supported by the Swiss National Science Foundation and by the NCCR
program \textit{Materials with Novel Electronic Properties}
(MaNEP) sponsored by the Swiss National Science Foundation.


%
\end{document}